\documentclass[12pt]{iopart}

\usepackage{iopams}
\usepackage{graphicx}

\begin{document}

\title{Colloidal suspensions in modulated light fields}

\author{M C Jenkins and S U Egelhaaf}

\address{Condensed Matter Physics Laboratory, Lehrstuhl f\"{u}r
Physik der weichen Materie, Heinrich-Heine-Universit\"{a}t
D\"{u}sseldorf, Universit\"{a}tstra{\ss}e 1, D-40225 D\"{u}sseldorf,
Germany.} \ead{matthew.jenkins@uni-duesseldorf.de}

\begin{abstract}
Periodically-modulated potentials in the form of light fields have
previously been applied to induce reversible phase transitions in
dilute colloidal systems with long-range interactions.  Here we
investigate whether similar transitions can be induced in very dense
systems, where interparticle contacts are important.  Using
microscopy we show that particles in such systems are indeed
strongly affected by modulated potentials. We discuss technical
aspects relevant to generating the light-induced potentials and to
imaging simultaneously the particles.  We also consider what happens
when the particle size is comparable with the modulation wavelength.
The effects of selected modulation wavelengths as well as pure
radiation pressure are illustrated.
\end{abstract}

\submitto{\JPCM}
\maketitle

\section{Introduction}
One of the defining features of soft matter systems, their
intermediate, or mesoscopic, lengthscale, places them at a
remarkable confluence of physical properties, where they are
susceptible to influence by ``everyday'' external forces. Not only
are they ``soft'' in their response to applied shear, and their
constituents sufficiently buoyant to display an interesting
interplay between gravity and thermal (Brownian) motion
\cite{Haw02}, colloids are also affected by the radiation
pressure--ordinarily thought of as negligible in laboratory
experiments--arising as they scatter light which impinges on them.

The coincidence of the magnitude of these forces (Brownian,
gravitational, and that arising from scattered EM radiation) in
these systems, set fundamentally by the relative magnitudes of
Boltzmann's constant $\mathrm{k_b}$, the gravitational constant
$\mathrm{G}$, the speed of light $\mathrm{c}$, and Planck's constant
$\mathrm{h}$, is compounded by a coincidence of the particle size
with the wavelength of visible light.  The remarkable consequence is
that we are able to study, in real space, Brownian particles under
the influence of light-induced potentials. Ashkin is credited with
first realising this possibility \cite{Ashkin70, Ashkin71, Ashkin74,
Ashkin80, Smith81, Ashkin82}, leading ultimately to his development
of optical tweezers \cite{Ashkin86}. These are now a widely-used
tool in physics \cite{Molloy02} and biology
\cite{Svoboda94,Sheetz98,Greulich99}. Studies using optical tweezers
generally consider very deep potential minima, where objects are
tightly trapped; the present work considers in addition the case
where potentials do not necessarily irreversibly capture particles
but do bias them towards certain locations.

Colloids have found favour as models of atomic systems, among their
advantages being the possibility of imaging directly large colloids
with light microscopy, the relatively slow speed at which they
diffuse, and the relative ease with which interactions can be
tailored \cite{PuseyLesHouches, Poon02}. In the apparently simple
case of hard-sphere interactions, colloidal systems show the
predicted entropy-driven fluid-solid transition at high density
\cite{Pusey86} and, at higher densities still, a glass transition
instead of the expected crystalline equilibrium state, again in
analogy with atomic and molecular glass-forming systems
\cite{Pusey86, Pusey87, vanMegen93, Weeks00, Ferrer98}. Similar
behaviour has been seen in other systems, for example charged
particles interacting via long-ranged electrostatic repulsion
\cite{Haertl95}.

Fluid-solid transitions are ubiquitous in nature, and are frequently
the expected equilibrium state.  Often, solid phases occur upon
cooling, whereupon distinct density modulations are enhanced,
resulting in translational and orientation order.  Periodically
modulated potentials can have a similar effect, giving rise to a
controllable phase transition effected at will by application and
removal of an external potential.  Chowdhury was the first to
demonstrate this, in two dimensions, in so-called laser-induced
freezing (LIF) experiments \cite{ChowdhuryThesis, Chowdhury85}. They
were followed by further experimental \cite{Ackerson87, Loudiyi92,
Loudiyi92b, Wei98b} and theoretical \cite{Xu86, Barrat90,
Chakrabarti94, Sood96} confirmation. In LIF, a
one-dimensionally-modulated potential gives rise to a
two-dimensionally-modulated crystal. The registration of adjacent
rows is mediated by Brownian motion perpendicular to the direction
of the potential, which the particles continue to undergo, although
they sit on average at the minima of the potential. For deeper
potentials, the lateral motion of the particles becomes sufficiently
restricted that the system is only modulated in one dimension, but
remains liquid-like along the direction of the fringes;
laser-induced melting (LIM) occurs \cite{Chakrabarti95, Wei98,
Bechinger00, Bechinger01, Bechinger01b, Bechinger02, Strepp01,
Strepp02, Strepp02b}.
There are many open questions relating to
light-induced phase transitions, ranging from the kinetics of LIF
and LIM, to as yet untested predictions for systems with more
complex inter-particle potentials \cite{Gotze03}, time-varying
potentials \cite{Rex05}, and binary mixtures \cite{Franzrahe07}.

LIF and LIM both represent transitions between equilibrium states,
brought about by the application of external, laser-induced
potentials.  In all of these cases, removal of the potential leads
to a re-establishment of the initial state. In contrast to
reversible light-induced transitions, there is the possibility that
externally-applied modulated potentials may induce expected but
unobserved transitions to ordered states. A prime example is the
hard-sphere glass transition described above, though other glass and
gel transitions may be relevant.  Dynamical arrest due to
co-operative effects such as caging is frequently offered as a
possible explanation for the glass transition, e.g. \cite{Weeks00,
Pham02, Goetze92}. It is known that shear can lead to ordering in
glasses and gels \cite{Ackerson88, Haw98, Haw98b, Vermant05,
Smith07}, though the underlying mechanism is still far from
understood \cite{Besseling06, Koumakis08}. (Note that shear can also
lead to melting of crystals \cite{Ackerson81, Stevens91}.)  If the
emerging picture of co-operative arrest is accurate, and given the
above evidence of external field-induced ordering, it seems
plausible that periodic light fields might also induce ordering in
very dense systems. Biroli {\em et al.} have developed an
inhomogeneous mode-coupling theory (MCT) which predicts the response
of a supercooled liquid's dynamical structure factor when exposed to
a static inhomogeneous potential \cite{Biroli06}. Agreement of these
calculations with experimental results would represent a powerful
test for MCT. 

This paper describes a first investigation into whether periodically
modulated light fields can indeed induce an effect in very dense
colloidal systems.  In a few initial steps towards the grand goals
mentioned above, we present some technical aspects relating to the
experimental realisation of an apparatus which permits exposure of a
sample to the light fields and its simultaneous observation. Our
first results with this apparatus show that even the most dense
samples do show a clear rearrangement under the influence of an
applied light field.


\section{Light as an external potential for colloidal particles}

Light forces acting on small particles have been described in
considerable detail, both from a ray-optics and a Rayleigh
perspective. The necessary basic physics is well established
\cite{Gordon73, Harada96, Tlusty98, Ashkin92}.  We outline the two
features most important to the present work, the so-called
scattering force $F_{\mathrm{scat}}$, which acts in the direction of
beam propagation, and the gradient force $F_{\mathrm{grad}}$, which
depends on the shape of the light intensity distribution.

\subsection{Scattering force $F_{\mathrm{scat}}$}

Historically, the scattering force preceded full optical trapping in
the form of Ashkin's levitation experiments \cite{Ashkin70,
Ashkin71, Ashkin74}.  It arises as a result of momentum transfer to
a particle from incident photons as they are scattered from it.

A photon of light with frequency $\nu$, wavelength $\lambda$ and
phase velocity $v_P$ carries energy
$E_{\mathrm{phot}}=\mathrm{h}\nu=\mathrm{h}v_P/\lambda$ and momentum
$p = \mathrm{h}/\lambda = E_{\mathrm{phot}}/v_P$.  In absorbing a
photon, an object therefore experiences a force $F_{\mathrm{scat}} =
\partial p/\partial t = \partial \left(E_{\mathrm{phot}}/v_P\right)/ \partial
t=P/v_P=nP/\mathrm{c}$, where $P$ is the power of the photon source.
The last equality follows since the phase velocity of light in a
material of refractive index $n$, $v_P=\mathrm{c}/n$, where
$\mathrm{c}$ is the speed of light {\em in vacuo} \cite{Ashkin86,
Molloy02}\footnote{We have assumed the Minkowski form, which differs
from the Abraham form by a factor $n^2$. Although this is an
important unresolved theoretical issue \cite{Leonhardt06}, this
subtlety does not concern us unduly, since optical forces can be
calibrated experimentally (and $n^2{\sim}1$).}. In our case, photons
are scattered rather than absorbed and therefore transfer only a
portion of their momentum to the particle; in the case of dielectric
spheres, a prefactor $q\sim0.1$ is customarily assumed
\cite{Ashkin70, Molloy02}.

In our experiments, we have calculated $F_{\mathrm{scat}}$ to be on
the order of a tenth of the particles' own weight for each $mW$ of
applied laser power (as measured at the laser, as always in this
article);
in a typical experiment this corresponds to around a tenfold
increase in the particles' effective buoyant mass.

\subsection{Gradient force $F_{\mathrm{grad}}$}

The gradient force $F_{\mathrm{grad}}$ allows a single beam, suitably
focused, to act as an optical trap.
Its origin is intuitively clear in the ray optics formulation
\cite{Molloy02, Ashkin92}, although since we ultimately consider the
sum of forces over all infinitesimal volume elements of finite-sized
spheres, we discuss the origin of the gradient force as it applies
to Rayleigh particles.  In this regime, the electric field over a
particle is approximately uniform, and induces a dipole moment
$\mathbf{p}=\alpha\mathbf{E}$, where $\alpha$ is the polarisability
of the (dielectric) particle given by the Clausius-Mosotti relation
\cite{JacksonBook}:
\[
\alpha=n_\mathrm{s}^2\left( \frac{n^2-1}{n^2+2} \right)a^3,
\]
with $n=n_\mathrm{c}/n_\mathrm{s}$ the ratio of the refractive
indices of the colloidal particle, $n_\mathrm{c}$, to the
surrounding medium (solvent), $n_\mathrm{s}$, and $a$ is the
particle's radius. Having acquired a dipole moment, the particle
experiences a Lorentz force $\mathbf{f}_\mathrm{L} =
(\mathbf{p}\cdot \nabla) \mathbf{E} + (1/\mathrm{c}) d\mathbf{p}/dt
\times\mathbf{B}$.  This can be re-written \cite{Gordon73}:
\[
\mathbf{f}_\mathrm{L} = \alpha\left(\nabla(\frac{1}{2}E^2)  +
\frac{1}{\mathrm{c}}\frac{\partial}{\partial
t}(\mathbf{E}\times\mathbf{B})\right).
\]
The second term is the scattering force $F_{\mathrm{scat}}$
discussed above, and is directed along the direction of propagation.
The first term, $F_{\mathrm{grad}} = (\alpha/2) \nabla E^2$, is the
gradient force and expresses that in a non-uniform electric field, a
particle will move towards regions of higher electric field if
$n_\mathrm{c}{>}n_\mathrm{s}$, and vice versa. This indicates that
the modulated potentials we seek can be realised by
spatially-modulated electric fields.

\subsection{Modulated potentials from modulated light fields}
\begin{figure}[h]
\begin{center}
\mbox{
\includegraphics[angle=0, height=5cm, bb=0 0 300 300]{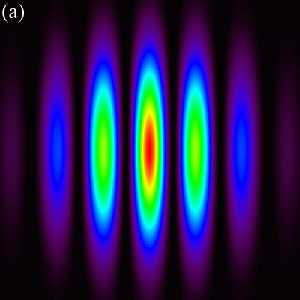} \quad
\includegraphics[angle=0, height=5cm, bb=0 0 187 310]{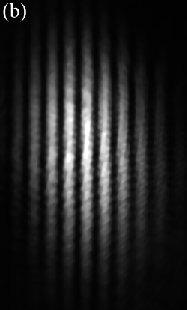} \quad
\includegraphics[angle=0, height=5cm, bb=0 0 300 300]{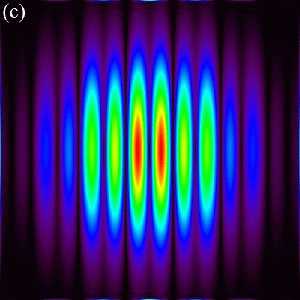}}
\caption{Fringe pattern calculated for typical experimental
parameters which result in a fringe spacing $d=9.29~\mu$m (a), an
observed fringe pattern (b), and the magnitude of the gradient force
calculated based on the fringe pattern shown on the left (c).
Parameters are (Section~\ref{apparatussection}): wavelength
$\lambda=532$nm, crossing angle $\theta=3.28^\circ$, beam radius at
the sample $R=1.22\lambda f /D=23.2\mu$m \cite{ChowdhuryThesis} for
a laser beam of diameter $D=2.80$~mm focused by a lens of focal
length $f=200$~mm. \label{fringedepiction}}
\end{center}
\end{figure}

\begin{figure}[h]
\begin{center}
\includegraphics[angle=0, width=12cm, bb=0 0 636 478]{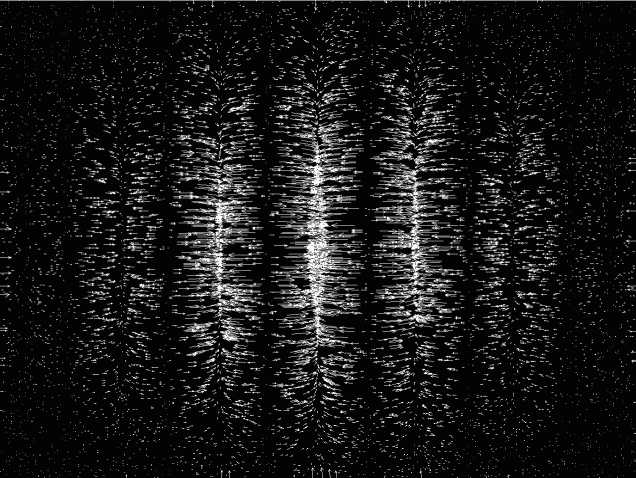}
\caption{A vector plot showing the position-dependent force
experienced by a particle, found by numerical differentiation
(Appendix 1) of the calculated pattern in Figure
\ref{fringedepiction} (left). \label{fringedepictionmoredetails}}
\end{center}
\end{figure}

One conceptually simple means of generating modulated light fields
is using a crossed beam experiment, as employed in previous studies
similar to ours \cite{Chowdhury85, Bechinger00}, and in thermal
diffusion forced Rayleigh scattering studies \cite{Wiegand04,
Koehler00}.  The experimental arrangement is also similar to the
so-called ``dual-scatter'' or ``dual-beam'' configurations used in
Laser Doppler Anemometry \cite{Brayton71, Durst76}. For two crossed
coherent TEM$_{00}$ mode (Gaussian profile) laser beams, the
resulting intensity profile is \cite{Loudiyi92}:
\begin{equation}
I(x) = 2\,I_0 \, \left\{ 1+\cos \left[
2kx\sin(\theta/2) \right] \right\} \, e^{-2x^2
\cos^2\left(\theta/2\right)/R^2}
\label{Ix}
\end{equation}
where $\theta$ is the beam crossing angle, $I_0$ the intensity of
each beam, $R$ the laser beam radius (the $e^{-2}$ point), and
$k=2\pi/\lambda$ the incident beam wavevector. The term in braces,
$\left\{ 1+\cos \left[ 2kx\sin(\theta/2) \right] \right\} = \left\{
1+\cos \left[ qx \right] \right\}$, is the sinusoidally-varying
interference pattern of fringe spacing $ d =
\lambda/\left(2\sin\left( \theta / 2 \right)\right)$ and fringe
wavevector  $q=2\pi /d$. The last term represents the overlying
Gaussian envelope due to the finite beam size. Figure
\ref{fringedepiction} compares a calculated fringe pattern (a) with
an experimentally observed pattern (b).  From the gradient, the
magnitude of the force can be calculated i.e.~$F_{\mathrm{grad}} =
(\alpha/2) \nabla E^2$ (c). These images illustrate that not only
are particles drawn to the fringes, but they are also confined by
the Gaussian envelope. Figure \ref{fringedepictionmoredetails}
illustrates this in the form of a vector field plot, which was
obtained by numerical differentiation (Appendix 1).

From the first term of the modulated light field, Equation \ref{Ix},
we obtain the oscillatory part of the potential: $ V(x) =
V_0\left\{1+\cos\left[qx\right]\right\}, $ where $V_0$ absorbs the
particles' polarisability and the laser beam intensity. Loudiyi {\em
et al.} additionally normalise this quantity by the thermal energy
\cite{Loudiyi92}.

\subsection{Effect of finite particle size}

Implicit in all of the previous section was that the potential acts
only at the centre of the particles, or alternatively that the
particle radius $a \ll d$. This is not satisfied in the present
work. For finite particle size, integration over infinitesimal
volume elements of the particle results in a modified potential
\cite{Loudiyi92}:
\begin{equation}
V(x) = V_0\left\{1 + 3\frac{j_1(qa)}{qa }\cos
\left[qx\right]\right\} \; , \label{finitesizeeqn}
\end{equation}
where $x$ is the position of the particle centre and $j_1$ the first
order spherical Bessel function.

\begin{figure}[h]
\begin{center}
\includegraphics[angle=0, width=9cm, clip, bb=0 0 698 488]{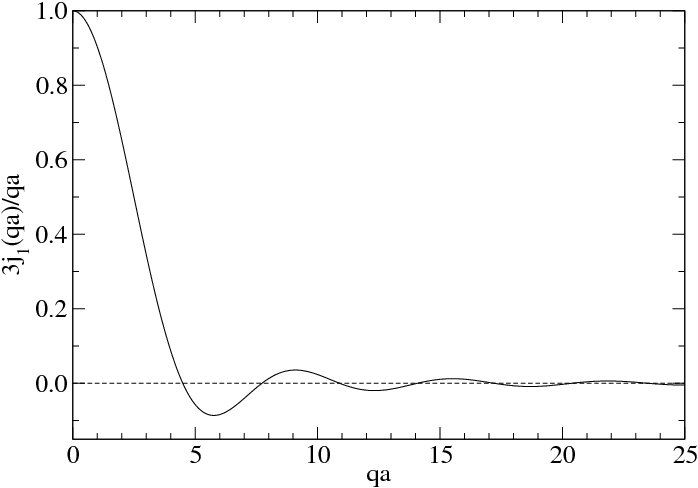}
\caption{The factor $3{j_1(qa)}/(qa)$ arising from the finite
particle radius $a$ (Equation \ref{finitesizeeqn}).
\label{finiteparticlesize}}
\end{center}
\end{figure}

The finite size of the particles leads to the additional factor
$3j_1(qa)/(qa)$ (Figure \ref{finiteparticlesize}). This factor
resembles the form factor of a sphere as it occurs in scattering
experiments \cite{Lindner02}. In a scattering experiment, the sample
is illuminated by a beam and the intensity scattered under a
scattering angle $\theta$ or, equivalently, a scattering vector $q$
is determined. This implies an identical geometry with the incident
and scattered beam here represented by the two crossing beams.

The additional factor demonstrates several interesting features
\cite{Chowdhury91}. First, for large fringe spacing ($qa\simeq0$),
the particles behave like point particles. Conversely, for very
small fringe spacing ($qa{\to}\infty$), the effective potential is
averaged to zero.  Interestingly, there are fringe spacings where
the sign of the potential is reversed, indicating that spherical
particles can be either drawn into the fringes or repelled from
them, depending on $qa$. As long as the fringe spacing is greater
than $d=0.699\times 2a$ (corresponding to $qa = 4.493$, the first
root of $j_1(qa)$), the factor $3{j_1(qa)}/(qa) > 0$. This means
that for fringe spacings larger than the particle diameter but still
finite (and indeed slightly smaller too), the behaviour is
qualitatively similar to that for point particles, albeit with
reduced potency.

\section{Experiment}

\subsection{Apparatus \label{apparatussection}}

\begin{figure}[h]
\begin{center}
\includegraphics[angle=0, width=15cm, clip, bb=0 0 990 536]{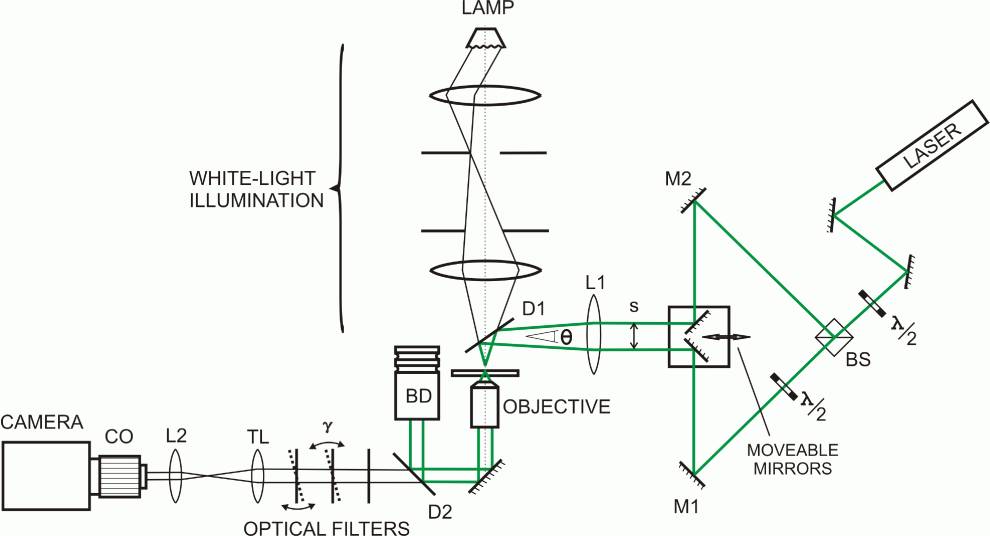}
\caption{Schematic diagram of the experimental arrangement used to
generate a modulated light field and simultaneously image the
sample.  Two coherent beams are created by beamsplitter BS, and
subsequently brought parallel by two mirrors (M1, M2) and a pair of
moveable mirrors. The position of the moveable mirrors determines
the beam separation $s$, and therefore the beam crossing angle
$\theta$ following focusing lens L1.  For observation, white
illuminating light is combined from above using dichroic mirror D1
and the sample imaged by an objective. The majority of the intense
laser light is deflected to beam dump BD using a second dichroic
mirror D2. A series of three optical filters, some rotatable, in the
objective infinity space allows for variable attenuation of the
remaining laser light and thus for an adjustment of the brightness
of the fringes to be viewed. Images are recorded, after suitable
magnification, using a digital camera.\label{apparatus}}
\end{center}
\end{figure}

Figure \ref{apparatus} shows the experimental arrangement used to
create the modulated potential and, at the same time, observe the
response of the sample. The modulated potential is created by
splitting a linearly-polarised laser beam (Coherent Verdi V5 with
$P=5$~W, $\lambda=532$~nm) and subsequently crossing the two beams.
The beam is split using a $50{:}50$ beamsplitter (BS) with a
preceding half-wave plate ($\lambda/2$) to adjust the polarisation
for optimum performance of the beamsplitter. The two beams are
brought parallel to one another by means of two mirrors (M1, M2) and
a moveable pair of mirrors. Translation of the moveable mirrors
adjusts the beam separation $s$ and, after the focusing lens (L1),
the crossing angle $\theta$ of the two beams. A half-wave plate
($\lambda/2$) in one of the beams allows rotation of the
polarisation of one beam with respect to the other, thereby
controlling the amplitude of the interference fringes whilst
maintaining a constant mean intensity and thus radiation pressure.

The introduction of the sample changes the crossing angle.  For a
typical sample cell (depth 170~$\mu$m) and fringe spacing
(7~$\mu$m), the angle of incidence is reduced from $\theta_i/2 =
\sin^{-1}\left({\lambda}/{2d}\right)=2.2^\circ$ to $\theta_r/2 =
\sin^{-1}\left({\sin(2.2^\circ)}/{1.33}\right)=1.6^\circ$,
corresponding to change in the focal position of
170~$\mu$m~${\times}\left({\tan(\theta_i/2)}/{\tan(\theta_r/2)} -1
\right)\simeq 64 \mu m$.  This is corrected by a linear translation
of the lens L1 (Figure \ref{apparatus}).  Note that despite the
change in the crossing angle, the fringe spacing remains unchanged
(since $\lambda$ also changes upon entering the new medium.)



Concurrently, the sample is imaged with a home-built inverted
bright-field microscope. K\"{o}hler illumination is provided from
above the sample.  The extremely long working distance condensing
lens provides sufficient space for a ``notch'' dichroic mirror (D1),
reflective in a narrow range around $\lambda=532$~nm but otherwise
transmitting in the visible. This dichroic mirror combines the
imaging and modulated light at the sample.  After the sample, a
standard high numerical aperture microscope objective (Nikon x100 PA
VC, NA=1.4) forms an image at infinity, before a tube lens (TL) and
subsequent telescope (L2, CO) adjust the magnification as
appropriate for the camera.  Additional optics can be introduced
straightforwardly into the (relatively long) so-called infinity
space behind the objective. We use this to separate the intense
laser light, damaging to the camera, from the white imaging light.
The bulk ($\simeq 98$\%) is deflected to a beam dump (BD) using a
second dichroic mirror (D2). As well as eliminating safely the
majority of the laser light, this light can be re-used, for example
by retro-reflection to achieve a counter-propagating arrangement
whereby $F_{\mathrm{scat}}$ can be reduced independently of the
laser intensity. Even after the second dichroic mirror, the
intensity of the laser light is far too high for the camera.  Three
additional optical filters are used to adjust the level of the
modulated light field while retaining most of the imaging light.
This permits simultaneous imaging of the sample and an appropriate
fraction of the light field.

\begin{figure}[h]
\begin{center}
\includegraphics[angle=0, width=9cm,clip, bb=0 0 699 491]{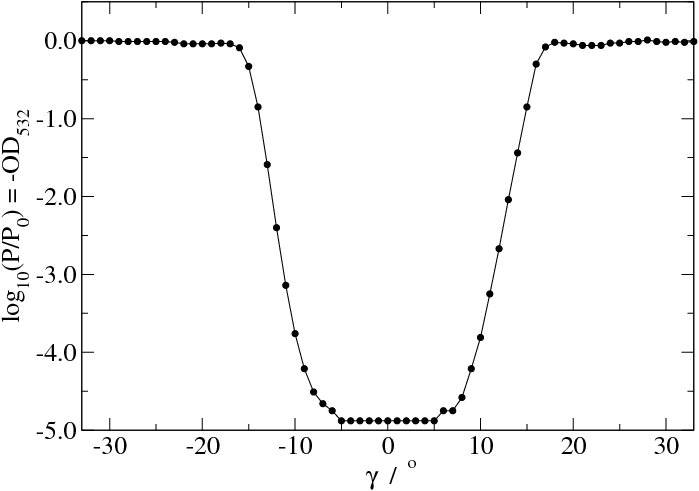}
\caption{Variation of the filter optical density at 532~nm,
$OD_{532}$, with the angle of incidence, $\gamma$.
 $P_0$ and $P$ are the measured power values before and after the filters, respectively. \label{filterODwithangle}}
\end{center}
\end{figure}

The intense laser light has to be reduced to about $10^{-9}$ --
$10^{-8}$~W at the camera, to obtain satisfactory images of the
fringes. The fringes become essentially invisible upon a further
reduction by a factor of about 100. Since the laser power varies
depending on the experiment, for optimum simultaneous imaging,
filters with a variable optical density at 532~nm, $OD_{532}$, are
desirable. This can be achieved for interference filters by changing
the angle of incidence, $\gamma$ (Figure \ref{apparatus}): the
filters have a sharp transmittance edge slightly above 532~nm which
shifts to lower wavelength as $\gamma$ is increased (similar to
notch filters~\cite{SemrockWeb}). Since the slope of the edge is
finite, this provides control over $OD_{532}$
(Figure~\ref{filterODwithangle}). When placed in the infinity space
of the microscope, these filters can be straightforwardly rotated to
allow the fringes to be imaged or not, as desired. The weak
dependence of $OD_{532}$ within $-5^\circ \lesssim \gamma \lesssim
5^\circ$ (Figure~\ref{filterODwithangle}) is important for imaging,
since this range is slightly larger than the divergence in the
infinity space of the microscope \footnote{Light from the focal
plane is focused at infinity, but, except for light originating from
the point on the optical axis, is nonetheless divergent. The focal
length of an objective is the microscope tube lens focal length
(here about 200~mm) divided by its magnification (here 100).
Together with the radius of the field of view (here about
125~$\mu$m), this results in a divergence in the infinity space of
the microscope of around $\tan^{-1}(125$~$\mu$m$ /
2$~mm$)\simeq3.6^\circ$.}. Nevertheless, for different parts of the
field of view, the effective $\gamma$ and thus $OD_{532}$ is
different, and hence an image of the fringes is no longer
quantitatively correct. A correct image can, however, be obtained
with $\gamma=0^\circ$, or by using neutral density filters (in which
case the white light is attenuated beyond usefulness).  The
bright-field images remain good since the transmittance of each
filter at wavelengths $\lambda \neq 532$~nm is $T \simeq 0.9$.

\subsection{Analysis}
\label{analysis} Having determined the particle coordinates
\cite{Crocker96, Jenkins08}, a range of parameters can be
calculated, for example the particle density $\phi$, the pair
correlation function $g(r)$, the mean coordination number $\langle z
\rangle$, the distribution of coordination numbers $p(z)$, and
bond-orientational order parameters, e.g. $\psi_6$
\cite{Villeneuve05}. Calibration of distances, necessary for
determining the fringe spacing as well as for structural analyses,
is performed using a high-resolution microscope test slide
(Richardson Test Slide, Model 80303) \cite{Jenkins08}.

\subsection{Samples}
We have used  polystyrene sulphate spheres of radius $a=2$~$\mu$m
(Interfacial Dynamics Corporation) suspended in water. The large
refractive index difference between particles ($n_\mathrm{c}=1.59$)
and water ($n_\mathrm{s}=1.33$) results in large optical gradient
forces, but the concomitant multiple scattering limits their use
effectively to a single layer, i.e.~two dimensions. These spheres
carry negative charges, which in the present study are screened by
high salt concentrations. We regard them as (almost) hard spheres,
which is supported by the observed distance of closest approach and
the shape of the pair correlation function (Section
\ref{twicenatural}). Though the salt concentration is high, it is
still low enough to avoid problems with coagulation.

Samples are prepared by pipetting a suitably-diluted homogenised
stock solution directly into the sample cell, which fills largely by
capillary action. The particles quickly sediment on the coverglass.
For dilute samples, this results in two-dimensional samples, while
at higher concentrations a few layers form (which can be reduced to
a single layer by application of radiation pressure,
Section~\ref{pureradpresssection}). The concentrations we refer to
in the following are the volume fractions $\phi$ of the initial,
homogenised, bulk solutions.  This is a nominal value; in the final
sample, inhomogeneities in the density may occur depending on the
settling process.

The sample cells consist of coverslips glued together, giving a
sample volume of about 20~mm~$\times$~3~mm~$\times$~170~$\mu$m
\cite{Jenkins08}. The sample only comes into contact with glass and
possibly the UV-cure glue used to seal the cells, whose effect is
assumed negligible. Since glass becomes negatively charged in the
presence of water, the particles are repelled from the surfaces of
the cell, and become attached only very occasionally
\cite{Prieve87,Prieve99}.

\section{Radiation pressure results \label{pureradpresssection}}

\begin{figure}[h]
\begin{center}
\includegraphics[angle=0, width=15cm, bb=0 0 654 386]{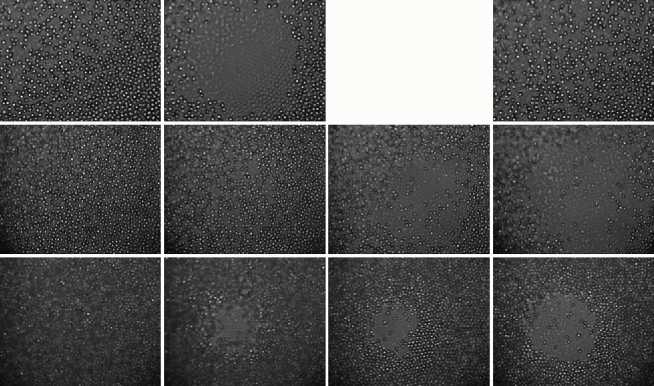}
\caption{The effect of pure radiation pressure as a function of
sample concentration (from top to bottom: initial homogenised volume
fractions $\phi=0.015$, 0.020, and 0.030) and laser intensity (left
to right: $P{=}$0.00~W, 0.10~W, 0.20~W, and 0.50~W). Images are
taken following 20 min of irradiation, except for the top right
image which shows the sample ($\phi=0.015$, $P$=0.10~W) 30 min after
the laser is turned off. Note that only the particles in the second
layer are in focus; they show a bright spot at their centre.
Particles in the first layer (most prominent in the image second
from left, top row) are out of focus but still clearly identifiable.
\label{pureradpressure}}
\end{center}
\end{figure}

We first investigate the effect of radiation pressure only as a
function of laser intensity and particle concentration (Figure
\ref{pureradpressure}). The radiation pressure is applied by turning
the second half-wave plate ($\lambda/2$) until minimum contrast is
achieved as judged from images of the interference patterns formed
using neutral density filters. The concentrations are chosen such
that a first layer of particles next to the coverslip (slightly out
of focus in Figure \ref{pureradpressure}) as well as an incomplete
second layer (in focus) are formed.

As the sample concentration increases (downwards in Figure
\ref{pureradpressure}), the density of the second layer increases.
The presence of a second layer does not imply that the maximum
possible density has been achieved in the first layer.  Indeed, this
is observed not to be the case: upon increasing the radiation
pressure, particles are pushed from the second layer into the first
layer (left to right). In each of these cases the radiation pressure
was applied for 20 min. In the least dense sample (top row), a laser
intensity $P=$0.10~W is already sufficient to insert all of the
particles into the first layer. (The two highest laser intensity
results for the lowest concentration are omitted in Figure
\ref{pureradpressure}.) This forms a dense hexagonally-close packed
(HCP) layer. With increasing density, a greater laser intensity is
required to insert all of the particles within the laser beam into
the first layer. Beyond a certain density, it is no longer possible
to insert all of the particles into the first layer, even for very
large radiation pressures. With increasing concentration, the area
with only a single layer gets smaller, corresponding to the Gaussian
profile of the laser beam and thus the applied radiation pressure.
This can also be seen with increasing laser intensity.

The top rightmost image shows the least concentrated sample 30 min
after the field is removed. (Similar relaxation behaviour is
observed in all samples.) The second layer has become repopulated.
This indicates that the osmotic pressure experienced within the
highly concentrated first layer is sufficient to cause particles to
``pop up'' into the second layer. This also implies that the
inability of particles to enter the first layer under their own
weight cannot be explained by pure geometrical frustration.
We have observed that ``popping up'' occurs with a characteristic
time of about 10~s. If this upward movement into the second layer is
thermally driven, i.e.~is a chance Brownian excursion (and a return
to the first layer is hindered by particle rearrangements within the
first layer), its timescale should be given by Kramer's escape time
with a ramp potential of depth $U_0$ and extent $2a$ representing
the gravitational potential \cite{Smith07, Kramers40}:
\[
\tau = \frac{1}{D_\mathrm{s}}\int_0^{2a} \mathrm{d}x' e^{\beta
U(x')} \int_{-\infty}^{x'}\mathrm{d}xe^{-\beta U(x)} =
\frac{(2a)^{2}}{D_\mathrm{s}} \frac{e^{-\beta U_0}-(1-\beta
U_0)}{(\beta U_0)^2},
\]
where $\beta=1/\mathrm{k_B}T$ and $D_\mathrm{s}$ is the
self-diffusion constant for a free particle. Using appropriate
parameters gives $\tau\sim 10^6$~s~$\gg 10$~s, indicating the
presence of a driving force, namely the osmotic pressure.

This seems plausible in light of the established connection between
the statistical geometry of hard spheres and their thermodynamic
properties \cite{Dullens05, Dullens05b, Widom63}.  These references
suggest that insertion of particles into a disordered layer by the
application of radiation pressure should, as in the Widom insertion
method, permit study of the thermodynamic properties of particles in
the first layer.

\section{Modulated potential results}

Having established a two-dimensional sample, we now introduce a
modulated potential. In the present study, the modulation is always
as great as possible at the specified laser intensity, i.e.~both
beams have the same direction of polarisation.   In each experiment,
we prepare a dense hexagonally-close packed (HCP) layer using
radiation pressure (at the indicated power) before the half-wave
plate is rotated to ``turn on'' the modulation. The parameters we
vary are the fringe separation and the amplitude of the modulation.
The HCP symmetry suggests a few fringe spacings $d$
(Figure~\ref{doublebeamspacingpossibilities}A); here we investigate
$d=\sqrt{3}\,a$ (Section~\ref{natural}) and $d=2\sqrt{3}\,a$
(Section~\ref{twicenatural})

\subsection{Natural fringe spacing}
\label{natural}


For a modulated potential of fringe spacing $d=\sqrt{3}\,a$, it is
possible for all particles forming an HCP layer to lie at the
potential minimum.  We thus consider this a natural fringe spacing.

\begin{figure}[h]
\begin{center}
\mbox{
\includegraphics[angle=0,width=7cm, bb=0 0 413 281]{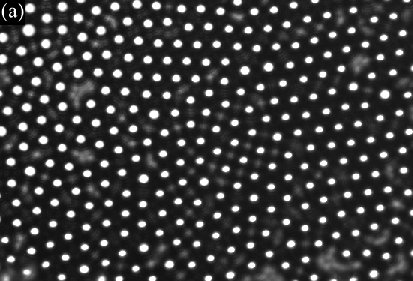}
\includegraphics[angle=0,width=7cm, bb=0 0 413 281]{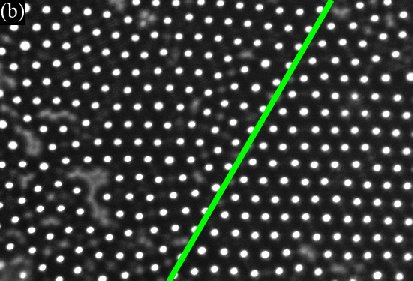}}\\
\caption{Sample (initial concentration $\phi=0.020$) after 30~min of
pure radiation pressure (a) followed by a further 30~min with a
modulated potential of wavelength $d=\sqrt{3}\,a$ (b). The laser
intensity in both cases is $P=$0.5~W, and the superimposed line
indicates the approximate fringe
direction.\label{singlebeamspacing}}
\end{center}
\end{figure}

\begin{figure}[h]
\begin{center}
\includegraphics[angle=0,width=9cm,clip, bb=0 0 702 491]{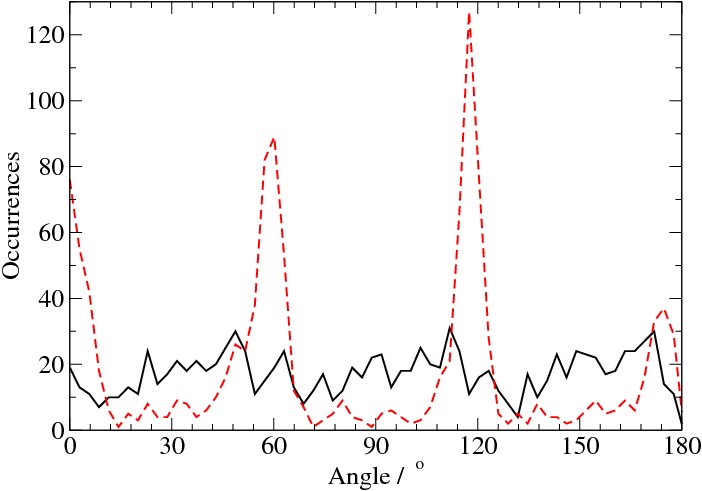}
\caption{Distribution of the nearest-neighbour bond direction for a
sample (initial concentration $\phi=0.020$) after 30~min of pure
radiation pressure (solid black line) followed by a further 30~min
with a modulated potential of wavelength $d=\sqrt{3}\,a$ (dashed red
line). The laser intensity is $P=$0.5~W.\label{singlebeamspacing2}}
\end{center}
\end{figure}

Pure radiation pressure (a single beam of $P=$0.50~W for 30~min)
leads to randomly-oriented crystallites (Figure
\ref{singlebeamspacing}, left). After exposure to the modulated
potential ($P=$0.50~W for 30~min), the crystallites have rotated and
consolidated to a near-perfect crystal with a clear direction
aligned with the fringes (right). This is also reflected in the
distribution of the nearest-neighbour bond direction, which shows
three strong peaks separated by $60^\circ$ (Figure
\ref{singlebeamspacing2}).

The crystallites thus seem to be able to rearrange despite the high
density. It is interesting to investigate exactly how this process
occurs. One observation is that as part of a crystallite rotates,
the total energy in the light field does not decrease monotonically
until the particles are aligned with the field.  At some angles
$\psi$ between the light field and the crystal orientation,
relatively many particles are near to the potential minima. When the
particles are aligned with the field ($\psi {=} 0^\circ$), all of
the particle centres occupy a minimum. For angles $\psi {\simeq}
18^\circ$, $31^\circ$, and $42^\circ$, there are only around $40$\%,
$60$\%, and $40$\% of the particles in the minimum respectively,
whereas for in-between angles there are far fewer. Supposing a large
crystallite were to rotate towards the global minimum, therefore, it
may do so at varying speed, perhaps even pausing at these
intermediate metastable orientations depending on the amplitude of
the field.

\subsection{Twice natural fringe spacing}
\label{twicenatural}

\begin{figure}[h]
\begin{center}
\includegraphics[angle=0,height=6cm, bb=0 0 205 265]{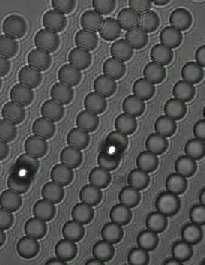}
\caption{A dilute sample exposed to a modulated light field
($P{=}$0.50W) with spacing $d=2\sqrt{3}\,a$.
\label{doublebeamspacingexpected}}
\end{center}
\end{figure}

We now consider a fringe spacing $d=2\sqrt{3}\,a$ corresponding to
twice the spacing between two rows of an HCP layer. For sufficiently
dilute samples, the particles align along the fringes (Figure
\ref{doublebeamspacingexpected}).

\begin{figure}[h]
\begin{center}
\mbox{
\includegraphics[angle=0,height=5.5cm, width=7.3cm, bb=0 0 480 360]{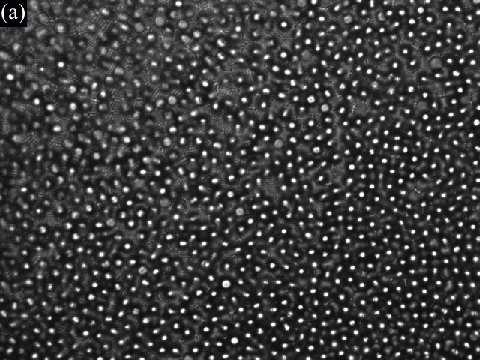}
\quad
\includegraphics[angle=0,height=5.5cm, width=7.3cm, bb=0 0 480 360]{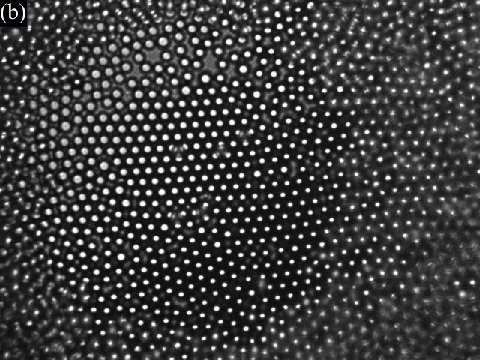}}
\mbox{
\includegraphics[angle=0,height=5.5cm, width=7.3cm,clip, bb=0 0 480 360]{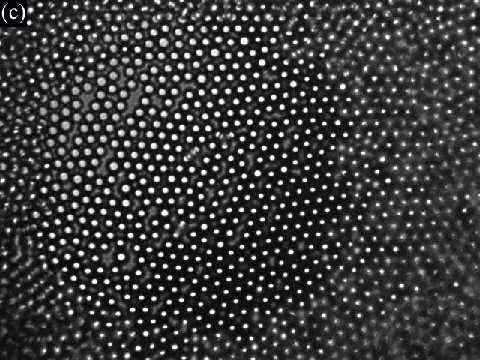}
\quad
\includegraphics[angle=0,height=5.5cm, width=7.3cm,clip, bb=0 0 480 360]{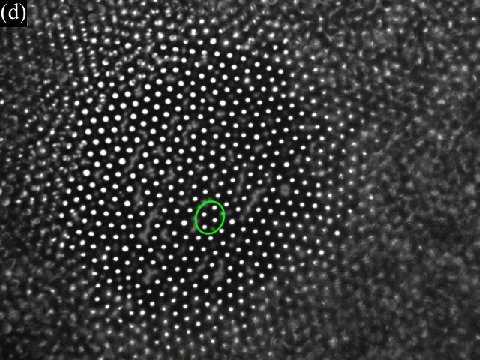}}
\caption{Micrographs of a sample (initial concentration
$\phi=0.020$) before irradiation (a), following 1~h of radiation
pressure at a laser intensity of $P=$0.40~W (b), and 100~s (c) and
about 5~h (d) after the introduction of fringes with spacing
$d=2\sqrt{3}\,a$. \label{doublebeamspacingexperiment}}
\end{center}
\end{figure}

\begin{figure}[h]
\begin{center}
\includegraphics[angle=0,width=15cm,bb=0 0 987 401]{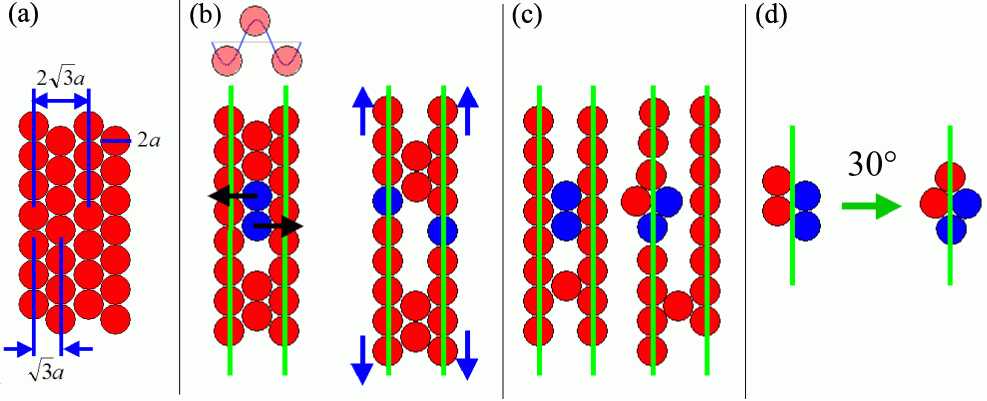}
\caption{(a) Hexagonally-close packed (HCP) layer of particles with
radius $a$ and inter-layer spacing $\sqrt{3}a$.  (b) Particles
located along the fringes (minima) are stable while those at the
maxima (indicated in dark blue) are metastable and, as a result of
fluctuations, experience a force toward the fringes. They can join a
minimum if particles which are already present in the minimum can
advance along it (blue arrows). (c) When this is hindered, the
particles can locally rearrange (e.g., rotate by $30^\circ$) to
adopt more favourable structures.
\label{doublebeamspacingpossibilities}}
\end{center}
\end{figure}

In dense samples, more complex structures develop (Figure
\ref{doublebeamspacingexperiment}). The initially disordered sample
(top left) develops randomly-oriented crystallites following the
application of radiation pressure (1~h of $P=$0.40~W, top right), as
described previously (Section~\ref{pureradpresssection}). Relatively
soon after a modulation of wavelength $d=2\sqrt{3}\,a$ is introduced
(100~s, still with $P=$0.40~W, bottom left), the sample is altered,
with the emergence of voids which run broadly in the direction of
the fringes. After substantially more time (about 5~h, bottom
right), the field has caused significant structural rearrangement.
In time-lapse movies of images, groups of clusters can be seen
moving co-operatively, leading to arrangements along the potential
minima. In particular, the motif highlighted in Figure
\ref{doublebeamspacingexperiment} (bottom right) occurs frequently,
with an orientation relative to the fringes as indicated in Figure
\ref{doublebeamspacingpossibilities}d. This rotation is
understandable on energetic grounds, which we discuss further below.
Other samples show similar behaviour.

\begin{figure}[ht]
\begin{center}
\includegraphics[angle=0,width=6cm,bb=0 0 740 2540]{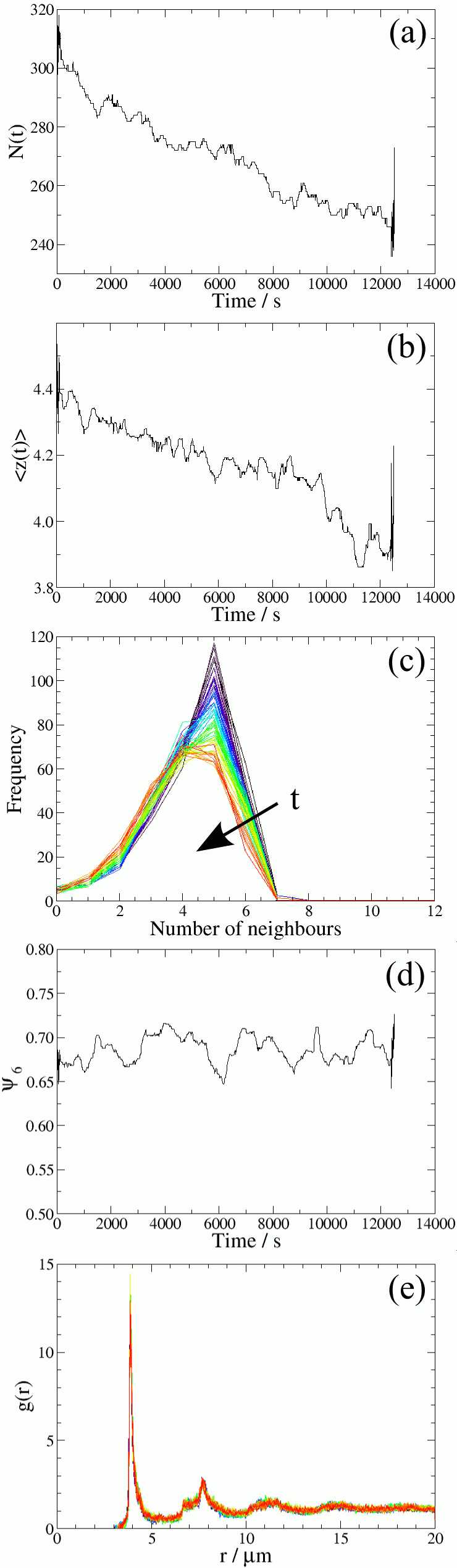}
\caption{Effect of a modulated potential on the evolution of
different parameters. Shown are the time dependence of (a) the
number of particles $N(t)$ within the observation region, (b) the
average number of neighbours $\langle z(t) \rangle$, (c) the
distribution of the number of neighbours $p(z,t)$ with time
(direction of increasing time indicated by arrow), (d) the
bond-orientational order parameter $\psi_6(t)$, and (e) the pair
correlation function $g(r,t)$. \label{doublebeamspacinganalysis}}
\end{center}
\end{figure}

The structural evolution of the sample has been investigated more
quantitatively by following the rearrangements induced by a
modulated potential (Figure \ref{doublebeamspacinganalysis}). We
determined the positions of particles which were located in a
rectangular region within the single-layer region and thus under the
influence of the modulated potential. Over the course of the whole
period, the number of particles $N(t)$ within the observation region
and thus the particle density steadily decreases (a). In addition, a
particle's average number of neighbours $\langle z(t) \rangle$ drops
from around 4.4 to 3.9 after 3.5~h (b). This is also reflected in
the distribution of the number of neighbours $p(z,t)$ (c), which
indicates an increasing probability of weakly connected particles,
consistent with the appearance of voids along the fringes.  Although
the number of neighbours decreases, the bond-orientational order
parameter $\psi_6$ (Section~\ref{analysis}) does not change
significantly over the course of the experiment (d).  This indicates
that those particles which remain bonded do so in a morphologically
similar way.  This is supported by the fact that the pair
correlation function $g(r)$ is also essentially unaffected
throughout the experiment (e).

How can we understand these observations?  In dilute samples, all of
the particles can be arranged in the potential minima. For the dense
samples, half of the particles can still lie along the minima
(Figure \ref{doublebeamspacingpossibilities}a)
(Section~\ref{natural}), but the remaining particles are forced to
lie between the fringes and thus at the maximum of the potential.
While the intensity gradient and hence the force is zero at the
maximum, this arrangement is metastable, with very small
fluctuations inevitably resulting in large gradient forces.  These
forces attempt to insert particles into the minima, i.e.~the fringes
(dark blue particles in Figure
\ref{doublebeamspacingpossibilities}b), and in so doing push other
particles along the fringes (blue arrows). This is achieved without
penalty, provided the density at the end of the fringe is suitably
low. When the density of the sample is large, there is a significant
osmotic penalty associated with pushing particles along the fringes
and into the bulk.  A balance must be struck between the optical
gradient force and the osmotic force, which are opposed in their
preference for density modulations.  This explains why the expected
modulations in density are observed at low concentrations (Figure
\ref{doublebeamspacingexpected}), but not at very high
concentration.  At high densities, the system aims to accommodate as
many particles as possible within the fringe, without significant
extension along the fringe. We have observed structures which
achieve this; one example is that highlighted in Figure
\ref{doublebeamspacingexperiment} (bottom right) and explained in
Figure \ref{doublebeamspacingpossibilities}d. This rhombic ``motif''
represents a part of the crystal which, after rotation through
$30^\circ$, reaches an energetically advantageous state (which
depends on the precise details of the potential, see below) without
a large extension along the fringe direction. The rearrangement of
small crystalline parts leaves bond orientations unchanged,
consistent with the observed essentially constant $\psi_6$, as well
as leaving inter-particle distances largely unchanged. This latter
observation is consistent with our finding that $g(r)$ does not
change substantially.  What modest extension along the fringes there
is expels some particles, in agreement with the decrease in the
particle number $N(t)$ and in turn the mean number of neighbours
$\langle z(t)\rangle$.

These observations might have interesting consequences. First, if
what we observe are equilibrium structures, it is remarkable that
they form via small crystalline parts which are broken away and
simply reoriented with respect to the applied potential.  It is,
however, also conceivable that, due to the geometrical frustration
in a dense system, these co-operative motions are the only means by
which the system can rearrange.  In this case, the observed
structures would correspond to non-equilibrium states liable to
further evolution; indeed the evolution of particle number $N(t)$
and mean coordination number $\langle z(t) \rangle$ suggests that
the samples are still evolving (Figure
\ref{doublebeamspacinganalysis}a,b).  Whether equilibrium or not, it
is clear that the modulated potential has a profound effect even in
these dense samples.  Our experiments also suggest that at
intermediate (in the present context, though these are still
relatively very dense samples) concentrations, novel structures
might form due to the competition between the imposed potential
which favours density modulations, and the osmotic pressure of the
system which opposes them.

Which structure is energetically or kinetically preferable depends
on the shape of the potential.  For example, for a square-well
potential, the particles can, to some extent, move laterally within
the fringe without penalty. Depending on the potential width and
separation, a wealth of structures has been predicted
\cite{Harreis02}. Although in that study the colloids remain near to
one another due to mutual attraction rather than osmotic pressure
(as in our case), the effect is seemingly similar.  For a potential
with monotonically increasing curvature, e.g.~a quadratic potential,
it is advantageous to displace particles from the minimum as little
as possible; ``zig-zag'' lines are expected. In the present case,
however, the curvature of the potential is non-monotonic and it
seems reasonable that some particles maintain their position while
others are significantly displaced from the potential minimum.
 Together with the influence of the osmotic pressure due to the bulk
 sample, this energy-minimisation argument justifies the existence
 of the observed motifs.

\section{Conclusion}
We have described an apparatus used to expose a sample to
sinusoidally-varying light fields and simultaneously image the
sample. To demonstrate its capabilities, we have investigated the
response of colloidal particles to the modulated potentials which
arise from the light field. We have shown that these potentials
influence even samples dense enough that the dynamics of their
constituent particles are severely restricted.  Currently we are
further improving the apparatus by including a counter-propagating
beam which will allow us to control the modulated potential and
radiation pressure independently.  This will be achieved by
replacing the beam dump by a retro-reflector.

Densely-packed effectively two-dimensional samples have been
generated using radiation pressures of different intensity. The
behaviour of these samples upon exposure to modulated potentials has
been investigated for two different modulation wavelengths.  This
has revealed co-operative structural rearrangements and final
structures which seem to result from a competition between the
optical gradient force and the osmotic pressure of the bulk sample.
While theoretical predictions for a sinusoidal potential are
lacking, similar theoretical calculations suggest structures
comparable to those we have observed.

With this apparatus, we can now investigate different situations:
first, for disorder-to-order transitions, specific predictions exist
for binary hard disc mixtures under similar conditions to those
described here \cite{Franzrahe07}.  Second, disorder-to-disorder
transitions are expected for systems with attractive interactions
exposed to modulated potentials \cite{Gotze03}. Both these
transitions represent reversible transitions starting from
equilibrium states, in which the initial states are recovered on
removal of the modulated potential. In contrast, in a third
situation, high-density non-equilibrium systems, in particular
repulsive and attractive glasses, might undergo irreversible
transitions from their non-equilibrium state to an ordered
equilibrium state upon exposure to a modulated potential. In this
case, structural rearrangements lead to stable configurations that
persist even after removal of the external potential. In addition to
revealing new physics, this might also have implications for
material sciences.


\section{Acknowledgements}
We thank Hartmut L\"{o}wen, Wilson Poon and Richard Hanes for
helpful discussions.  We also thank J\"urgen Liebetrau for technical
assistance and Beate Moser for help in preparing the diagrams.  This
work was funded by the Deutsche Forschungsgemeinshaft (DFG) within
 the German-Dutch Collaborative Research Centre
 Sonderforschungsbereich-Transregio 6 (SFB-TR6), Project Section
 C7.

\section*{Appendix 1: Numerical differentiation of patterns \label{gradient}}
The calculated intensity profile was differentiated numerically using
the Sobel method \cite[\S7.1.3]{GonzalezWoods} to obtain an
approximation to the force field experienced by the particles. If
$f$ is the image, then the gradient of the image
\[
\nabla {\mathbf f} = \left[
\begin{array}{c}
G_x\\
G_y
\end{array}
\right] = \left[
\begin{array}{c}
\partial f / \partial x\\
\partial f / \partial y
\end{array}
\right],
\]
with magnitude $|\nabla{\mathbf f}|=(G_x^2 + G_y^2)^{1/2}$ and
direction $\varphi(x,y) = \tan^{-1}(G_y/G_x)$ is formed by
convolution of the image with the following kernels:

\[
G_x =
\begin{array}{|c|c|c|}
\hline -1&-2&-1\\
\hline 0&0&0\\
\hline 1&2&1\\\hline
\end{array} \qquad
G_y =
\begin{array}{|c|c|c|}
\hline -1&0&1\\
\hline -2&0&2\\
\hline -1&0&1\\\hline
\end{array}
\]

\section*{References}

\bibliographystyle{unsrt}
\bibliography{BibliographicDatabase_at11June08}

\begin{thebibliography}{10}

\bibitem{Haw02}
M.~D. Haw.
\newblock Colloidal suspensions, \uppercase{B}rownian motion, molecular
  reality: a short history.
\newblock {\em J. Phys.:Condens. Matter}, 14:7769--7779, 2002.

\bibitem{Ashkin70}
A.~Ashkin.
\newblock Acceleration and trapping of particles by radiation pressure.
\newblock {\em Phys. Rev. Lett.}, 24:156--159, 1970.

\bibitem{Ashkin71}
A.~Ashkin and J.~M. Dziedzic.
\newblock Optical levitation by radiation pressure.
\newblock {\em Appl. Phys. Lett.}, 19:283--285, 1971.

\bibitem{Ashkin74}
A.~Ashkin and J.~M. Dziedzic.
\newblock Stability of optical levitation by radiation pressure.
\newblock {\em Appl. Phys. Lett.}, 24:586--588, 1974.

\bibitem{Ashkin80}
A.~Ashkin.
\newblock Applications of laser radiation pressure.
\newblock {\em Science}, 210:1081--1088, 1980.

\bibitem{Smith81}
P.~W. Smith, A.~Ashkin, and W.~J. Tomlinson.
\newblock Four-wave mixing in an artificial \uppercase{K}err medium.
\newblock {\em Opt. Lett.}, 6:284--286, 1981.

\bibitem{Ashkin82}
A.~Ashkin, J.~M. Dziedzic, and P.~W. Smith.
\newblock Continuous-wave self-focusing and self-trapping of light in
  artificial \uppercase{K}err media.
\newblock {\em Opt. Lett.}, 7:276--278, 1982.

\bibitem{Ashkin86}
A.~Ashkin, J.~M. Dziedzic, J.~E. Bjorkholm, and S.~Chu.
\newblock Observation of a single-beam gradient force optical trap for
  dielectic particles.
\newblock {\em Opt. Lett.}, 11:288--290, 1986.

\bibitem{Molloy02}
J.~E. Molloy and M.~J. Padgett.
\newblock Lights, action: optical tweezers.
\newblock {\em Contemp. Phys.}, 43:241--258, 2002.

\bibitem{Svoboda94}
K.~Svoboda and S.~M. Block.
\newblock Biological applications of optical forces.
\newblock {\em Annu. Rev. Biophy. Biomol. Struct.}, 23:247--285, 1994.

\bibitem{Sheetz98}
M.~P. Sheetz.
\newblock {\em Laser Tweezers in Cell Biology (Methods in Cell Biology)}.
\newblock Academic Press, 1998.

\bibitem{Greulich99}
K.~O. Greulich.
\newblock {\em Micromanipulation by Light in Biology and Medicine}.
\newblock Springer, 1999.

\bibitem{PuseyLesHouches}
P.~N. Pusey.
\newblock {\em Liquids, Freezing and Glass Transition}, chapter 10. Colloidal
  Suspensions, pages 763--942.
\newblock Elsevier, Amsterdam, 1991.

\bibitem{Poon02}
W.~C.-K. Poon.
\newblock The physics of a model colloid-polymer mixture.
\newblock {\em J. Phys.:Cond. Mat.}, 14:R859--R880, 2002.

\bibitem{Pusey86}
P.~N. Pusey and W.~van Megan.
\newblock Phase behaviour of concentrated suspensions of nearly hard colloidal
  spheres.
\newblock {\em Nature}, 320:340--342, 1986.

\bibitem{Pusey87}
P.~N. Pusey and W.~van Megan.
\newblock Observation of a glass transition in suspensions of spherical
  colloidal particles.
\newblock {\em Phys. Rev. Lett.}, 59:2083--2086, 1987.

\bibitem{vanMegen93}
W.~van Megen and S.~M. Underwood.
\newblock Dynamic-light-scattering study of glasses of hard colloidal spheres.
\newblock {\em Phys. Rev. E}, 47:248--261, 1993.

\bibitem{Weeks00}
E.~R. Weeks, J.~C. Crocker, A.~C. Levitt, A.~B. Schofield, and D.~A. Weitz.
\newblock Three-dimensional direct imaging of structural relaxation near the
  colloidal glass transition.
\newblock {\em Science}, 287:627--631, 2000.

\bibitem{Ferrer98}
M.~L. Ferrer, C.~Lawrence, B.~G. Demirjian, D.~Kivelson, C.~Alba-Simionesco,
  and G.~Tarjus.
\newblock Supercooled liquids and the glass transition: \uppercase{T}emperature
  as the control variable.
\newblock {\em J. Chem. Phys.}, 109:8010--8015, 1998.

\bibitem{Haertl95}
W.~H\"artl, H.~Versmold, and X.~Zhang-Heider.
\newblock The glass transition of charged polymer colloids.
\newblock {\em J. Chem. Phys.}, 102:6613--6618, 1995.

\bibitem{ChowdhuryThesis}
A.~H. Chowdhury.
\newblock {\em Laser Induced Freezing}.
\newblock PhD thesis, Oklahoma State University, 1986.

\bibitem{Chowdhury85}
A.~Chowdhury and B.~J. Ackerson.
\newblock Laser-induced freezing.
\newblock {\em Phys. Rev. Lett.}, 55:833--837, 1985.

\bibitem{Ackerson87}
B.~J. Ackerson and A.~H. Chowdhury.
\newblock Radiation pressure as a technique for manipulating the particle order
  in colloidal suspensions.
\newblock {\em Faraday Discuss. Chem. Soc.}, 83:309--316, 1987.

\bibitem{Loudiyi92}
K.~Loudiyi and B.~J. Ackerson.
\newblock Direct observation of laser induced freezing.
\newblock {\em Physica A}, 184:1--25, 1992.

\bibitem{Loudiyi92b}
K.~Loudiyi and B.~J. Ackerson.
\newblock \uppercase{M}onte \uppercase{C}arlo simulation of laser induced
  freezing.
\newblock {\em Physica A}, 184:26--41, 1992.

\bibitem{Wei98b}
Q.-H. Wei, C.~Bechinger, D.~Rudhardt, and P.~Leiderer.
\newblock Structure of two-dimensional colloidal systems under the influence of
  an external modulated light field.
\newblock {\em Progr. Colloid Polym. Sci.}, 110:46--49, 1998.

\bibitem{Xu86}
H.~Xu and M.~Baus.
\newblock Freezing in the presence of a periodic external potential.
\newblock {\em Phys. Lett. A}, 117:127--131, 1986.

\bibitem{Barrat90}
J.~L. Barrat and H.~Xu.
\newblock The phase diagram of hard spheres in a periodic external potential.
\newblock {\em J. Phys.: Condens. Matter}, 2:9445--9450, 1990.

\bibitem{Chakrabarti94}
J.~Chakrabarti, H.~R. Krishnamurthy, and A.~K. Sood.
\newblock Density functional theory of laser-induced freezing in colloidal
  suspensions.
\newblock {\em Phys. Rev. Lett.}, 73:2923--2926, 1994.

\bibitem{Sood96}
A.~K. Sood.
\newblock Some novel states of colloidal matter: modulated liquid, modulated
  crystal and glass.
\newblock {\em Physica A}, 224:34--47, 1996.

\bibitem{Chakrabarti95}
J.~Chakrabarti, H.~R. Krishnamurthy, A.~K. Sood, and S.~Sengupta.
\newblock Reentrant melting in laser field modulated colloidal suspensions.
\newblock {\em Phys. Rev. Lett.}, 75:2232--2235, 1995.

\bibitem{Wei98}
Q.-H. Wei, C.~Bechinger, D.~Rudhardt, and P.~Leiderer.
\newblock Experimental study of laser-induced melting in two-dimensional
  colloids.
\newblock {\em Phys. Rev. Lett.}, 81:2606--2609, 1998.

\bibitem{Bechinger00}
C.~Bechinger, Q.~H. Wei, and P.~Leiderer.
\newblock Reentrant melting of two-dimensional colloidal systems.
\newblock {\em J. Phys.: Condens. Matter}, 12:A425--A430, 2000.

\bibitem{Bechinger01}
C.~Bechinger, M.~Brunner, and P.~Leiderer.
\newblock Phase behavior of two-dimensional colloidal systems in the presence
  of periodic light fields.
\newblock {\em Phys. Rev. Lett.}, 86:930--933, 2001.

\bibitem{Bechinger01b}
C.~Bechinger and E.~Frey.
\newblock Phase behaviour of colloids in confining geometry.
\newblock {\em J. Phys.:Condens. Matter}, 13:R321--R336, 2001.

\bibitem{Bechinger02}
C.~Bechinger.
\newblock Colloidal suspensions in confined geometries.
\newblock {\em Curr. Opin. Colloid Interface Sci.}, 7:204--209, 2002.

\bibitem{Strepp01}
W.~Strepp, S.~Sengupta, and P.~Nielaba.
\newblock Phase transitions of hard disks in external potentials: a
  \uppercase{M}onte \uppercase{C}arlo study.
\newblock {\em Phys. Rev. E}, 63:046106, 2001.

\bibitem{Strepp02}
W.~Strepp, S.~Sengupta, and P.~Nielaba.
\newblock Phase transitions of soft disks in external potentials: a
  \uppercase{M}onte \uppercase{C}arlo study.
\newblock {\em Phys. Rev. E}, 66:056109, 2002.

\bibitem{Strepp02b}
W.~Strepp, S.~Sengupta, M~Lohrer, and P.~Nielaba.
\newblock Phase transitions of hard and soft disks in external periodic
  potentials: a \uppercase{M}onte \uppercase{C}arlo study.
\newblock {\em Comput. Phys. Commun.}, 147:370--373, 2002.

\bibitem{Gotze03}
I.~O. G\"{o}tze, J.~M. Brader, M.~Schmidt, and H.~L\"{o}wen.
\newblock Laser-induced condensation in colloid-polymer mixtures.
\newblock {\em Mol. Phys.}, 101:1651--1658, 2003.

\bibitem{Rex05}
M.~Rex, H.~L\"{o}wen, and C.~N. Likos.
\newblock Soft colloids driven and sheared by traveling wave fields.
\newblock {\em Phys. Rev. E}, 72:021404, 2005.

\bibitem{Franzrahe07}
K.~Franzrahe and P.~Nielaba.
\newblock Entropy versus energy: The phase behavior of a hard-disk mixture in a
  periodic external potential.
\newblock {\em Phys. Rev. E}, 76:061503, 2007.

\bibitem{Pham02}
K.~N. Pham, A.~M. Puertas, J.~Bergenholtz, S.~U. Egelhaaf, A.~Moussa\"{\i}d,
  P.~N. Pusey, A.~B. Schofield, M.~E. Cates, M.~Fuchs, and W.~C.~K. Poon.
\newblock Multiple glassy states in a simple model system.
\newblock {\em Science}, 296:104--106, 2002.

\bibitem{Goetze92}
W.~G\"{o}tze and L.~Sjogren.
\newblock Relaxation processes in supercooled liquids.
\newblock {\em Rep. Prog. Phys.}, 55:241--376, 1992.

\bibitem{Ackerson88}
B.~J. Ackerson and P.~N. Pusey.
\newblock Shear-induced order in suspensions of hard spheres.
\newblock {\em Phys. Rev. Lett.}, 61:1033–1036, 1988.

\bibitem{Haw98}
M.~D. Haw, W.~C.~K. Poon, P.~N. Pusey, P.~Hebraud, and F.~Lequeux.
\newblock Colloidal glasses under shear strain.
\newblock {\em Phys. Rev. E}, 58:4673--4682, 1998.

\bibitem{Haw98b}
M.~D. Haw, W.~C.~K. Poon, and P.~N. Pusey.
\newblock Direct observation of oscillatory-shear-induced order in colloidal
  suspensions.
\newblock {\em Phys. Rev. E}, 57:6859--6864, 1998.

\bibitem{Vermant05}
J.~Vermant and M.~J. Solomon.
\newblock Flow-induced structure in colloidal suspensions.
\newblock {\em J. Phys.: Condens. Matter}, 17:R187--R216, 2005.

\bibitem{Smith07}
P.~A. Smith, G.~Petekidis, S.~U. Egelhaaf, and W.~C.~K. Poon.
\newblock Yielding and crystallization of colloidal gels under oscillatory
  shear.
\newblock {\em Phys. Rev. E}, 76:041402, 2007.

\bibitem{Besseling06}
R.~Besseling, E.~R. Weeks, A.~B. Schofield, and W.~C.~K. Poon.
\newblock Three-dimensional imaging of colloidal glasses under steady shear.
\newblock {\em Phys. Rev. Lett.}, 99:028301, 2007.

\bibitem{Koumakis08}
N.~Koumakis, A.~B. Schofield, and G.~Petekidis.
\newblock Effects of shear-induced crystallization on the rheology and ageing
  of hard sphere glasses.
\newblock {\em arxiv:0804.1218}, 2008.

\bibitem{Ackerson81}
B.~J. Ackerson and N.~A. Clark.
\newblock Shear-induced melting.
\newblock {\em Phys. Rev. Lett.}, 46:123--127, 1981.

\bibitem{Stevens91}
M.~J. Stevens, M.~O. Robbins, and J.~F. Belak.
\newblock Shear melting of colloids: a nonequilibrium phase diagram.
\newblock {\em Phys. Rev. Lett.}, 66:3004--3007, 1991.

\bibitem{Biroli06}
G.~Biroli, J.-P. Bouchaud, K.~Miyazaki, and D.~R. Reichman.
\newblock Inhomogeneous mode-coupling theory and growing dynamic length in
  supercooled liquids.
\newblock {\em Phys. Rev. Lett.}, 97:195701, 2006.

\bibitem{Gordon73}
J.~P. Gordon.
\newblock Radiation forces and momenta in dielectric media.
\newblock {\em Phys. Rev. A}, 8:14--21, 1973.

\bibitem{Harada96}
Y.~Harada and T.~Asakura.
\newblock Radiation forces on a dielectric sphere in the \uppercase{R}ayleigh
  scattering regime.
\newblock {\em Opt. Commun.}, 124:529--541, 1996.

\bibitem{Tlusty98}
T.~Tlusty, A.~Meller, and R.~Bar-Ziv.
\newblock Optical gradient forces of strongly localized fields.
\newblock {\em Phys. Rev. Lett.}, 81:1738--1741, 1998.

\bibitem{Ashkin92}
A.~Ashkin.
\newblock Forces of a single-beam gradient laser trap on a dielectric sphere in
  the ray optics regime.
\newblock {\em Biophys. J.}, 61:569--582, 1992.

\bibitem{Leonhardt06}
U.~Leonhardt.
\newblock Momentum in uncertain light.
\newblock {\em Nature}, 444:823--824, 2006.

\bibitem{JacksonBook}
J.~D. Jackson.
\newblock {\em Classical Electrodynamics}.
\newblock Wiley, 2$^{\mathrm{nd}}$ edition, 1975.

\bibitem{Wiegand04}
S.~Wiegand.
\newblock Thermal diffusion in liquid mixtures and polymer solutions.
\newblock {\em J. Phys.: Condens. Matter}, 16:R357--R379, 2004.

\bibitem{Koehler00}
W.~K\"{o}hler and R.~Sch\"{a}fer.
\newblock Polymer analysis by thermal-diffusion forced \uppercase{R}ayleigh
  scattering.
\newblock {\em Adv. Polym, Sci.}, 151:1--59, 2000.

\bibitem{Brayton71}
D.~B. Brayton and W.~H. Goethert.
\newblock A new dual-scatter laser \uppercase{D}oppler-shift velocity measuring
  technique.
\newblock {\em ISA Trans.}, 10:40--50, 1971.

\bibitem{Durst76}
F.~Durst, A.~Melling, and J.~H. Whitelaw.
\newblock {\em Principles and practice of laser-\uppercase{D}oppler
  anemometry}.
\newblock Academic Press, 1976.

\bibitem{Lindner02}
P.~Lindner and T.~Zemb (eds.).
\newblock {\em Neutrons, X-rays and Light: Scattering Methods Applied to Soft
  Condensed Matter}.
\newblock Elsevier, 2002.

\bibitem{Chowdhury91}
A.~H. Chowdhury, F.~K. Wood, and B.~J. Ackerson.
\newblock Transverse radiation pressure forces for finite sized colloidal
  particles.
\newblock {\em Opt. Commun.}, 86:547--554, 1991.

\bibitem{SemrockWeb}
Semrock Inc.
\newblock Notch filter spectra vs. angle of incidence.
\newblock http://www.semrock.com/Catalog/Notch\_SpectrumvsAOI.htm.
\newblock Current as of March 2008.

\bibitem{Crocker96}
J.~C. Crocker and D.~G. Grier.
\newblock Methods of digital video microscopy for colloidal studies.
\newblock {\em J. Colloid Interface Sci.}, 179:298--310, 1996.

\bibitem{Jenkins08}
M.~C. Jenkins and S.~U. Egelhaaf.
\newblock Confocal microscopy of colloidal particles: Towards reliable, optimum
  coordinates.
\newblock {\em Adv. Colloid Interface Sci.}, 136:65--92, 2008.

\bibitem{Villeneuve05}
V.~W.~A. de~Villeneuve, R.~P.~A. Dullens, D.~G. A.~L. Aarts, E.~Groeneveld,
  J.~H. Scherff, W.~K. Kegel, and H.~N.~W. Lekkerkerker.
\newblock Colloidal hard-sphere crystal growth frustrated by large spherical
  impurities.
\newblock {\em Science}, 309:1231--1233, 2005.

\bibitem{Prieve87}
D.~C. Prieve and F.~Loo.
\newblock Brownian motion of a hydrosol particle in a colloidal force field.
\newblock {\em Faraday Discuss. Chem. Soc.}, 83:297--307, 1987.

\bibitem{Prieve99}
D.~C. Prieve.
\newblock Measurement of colloidal forces with \uppercase{TIRM}.
\newblock {\em Adv. Colloid Interface Sci.}, 82:93--125, 1999.

\bibitem{Kramers40}
H.~A. Kramers.
\newblock Brownian motion in a field of force and the diffusion model of
  chemical reactions.
\newblock {\em Physica}, 7:284--304, 1940.

\bibitem{Dullens05}
R.~P.~A. Dullens, D.~G. A.~L. Aarts, and W.~K. Kegel.
\newblock Direct measurement of the free energy by optical microscopy.
\newblock {\em Proc. Nat. Acad. Sci. U.S.A.}, 103:529--531, 2006.

\bibitem{Dullens05b}
R.~P.~A. Dullens, D.~G. A.~L. Aarts, W.~K. Kegel, and H.~N.~W. Lekkerkerker.
\newblock The \uppercase{W}idom insertion method and ordering in small
  hard-sphere systems.
\newblock {\em Mol. Phys.}, 103:3195--3200, 2005.

\bibitem{Widom63}
B.~Widom.
\newblock Some topics in the theory of fluids.
\newblock {\em J. Chem. Phys.}, 39:2808--2812, 1963.

\bibitem{Harreis02}
H.~M. Harreis, M.~Schmidt, and H.~L\"{o}wen.
\newblock Decoration lattices of colloids adsorbed on stripe-patterned
  substrates.
\newblock {\em Phys. Rev. E}, 65:041602, 2002.

\bibitem{GonzalezWoods}
R.~C. Gonzalez and R.~E. Woods.
\newblock {\em Digital Image Processing}.
\newblock Addison-Wesley, 1992.

\end{thebibliography}

\end{document}